# Polarizable Mean-Field Model of Water for Biological Simulations with Amber and Charmm force fields

Igor V. Leontyev and Alexei A. Stuchebrukhov*

*Department of Chemistry, University of California Davis, One Shields Avenue, Davis, California 95616*

*E-mails: ileontyev@ucdavis.edu

## Abstract

Although a great number of computational models of water are available today, the majority of current biological simulations are done with simple models, such as TIP3P and SPC, developed almost thirty years ago and only slightly modified since then. The reason is that the non-polarizable force fields that are mostly used to describe proteins and other biological molecules are incompatible with more sophisticated modern polarizable models of water. The issue is electronic polarizability: in liquid state, in protein, and in vacuum the water molecule is polarized differently, and therefore has different properties; thus the only way to describe all these different media with the same model is to use a polarizable water model. However, to be compatible with the force field of the rest of the system, e.g. a protein, the latter should be polarizable as well. Here we describe a novel model of water that is in effect polarizable, and yet compatible with the standard non-polarizable force fields such as AMBER, CHARMM, GROMOS, OPLS, etc. Thus the model resolves the outstanding problem of incompatibility.

# 1. Introduction

In biological computer simulations [1-2], water plays as important a role as in real life. Water not only solvates the simulated molecules, but it also penetrates inside the proteins, becoming an integral part of the enzymes, it determines how molecules move, bind to each other, how they solvate charges, and screen electrostatic interactions.

Despite its obvious importance, today simulations of water in proteins are mostly done with the most primitive models, such as TIP3P or SPC, developed almost thirty years ago, and only slightly modified since then. The main drawback of these models is that they are non-polarizable. In liquid state, in protein, and in vacuum the water molecule is polarized differently, and therefore has different properties. The models such as TIP3P or similar, despite its name (T – for transferable), do not reflect this change when used for example inside or outside the protein, at interfaces, or for simulation of small clusters. To address this obvious problem, a number of polarizable models have been developed, however, they are incompatible[3] with the standard force fields such as AMBER[4-5], CHARMM[6], GROMOS[7], OPLS[8-9] etc. The reason is that such force fields are themselves non-polarizable. [For the development of fully polarizable models see Refs. [10-14]] The incompatibility issue arises due to differences in parameterization strategies of polarizable and non-polarizable force fields.

In general, the effects of electronic polarizability in empirical non-polarizable force fields are described implicitly in the form of the effective charges and other parameters; whereas polarizable water models describe electronic polarizability explicitly, and therefore do not match the nature of effective non-polarizable force fields. What is more important, the effective charges correspond to electronically screened interactions in the *condensed medium*, while fully polarizable models of water describe interactions in *vacuum*. (For example in CHARMM (ref [6]) an empirical factor of 1.16 is used in matching ab initio energy to reflect the combined effect of the condensed medium, while for polarizable moles this factor is exactly unity, as their goal is to exactly match ab initio and force field interactions in vacuum. Practically, as will be shown, the difference in electrostatic energy amounts to a factor of about 0.7.) It seems the incompatibility issue can be overcome only with the development of fully polarizable biomolecular force fields[10-11,14].

Here we describe an *effective* polarizable model that addresses the problem of incompatibility. The key ingredient of the model is that it separates the generic electronic screening effects of the condensed medium, which rather accurately can be described implicitly within the electronic continuum model [15-16], and the remaining polarization effects due to specific environment, i.e. local charges of the protein. The key idea is that the model treats the generic electronic screening effect of the condensed medium on the same footing as do the empirical non-polarizable force fields. The model is also utilizing the ideas of mean field theory and self-consistency, resulting in an effective polarizable TIP3P-like model that on the one hand treats electronic screening effects in the same way as all non-polarizable force fields, and on the other hand, adjusts the water dipole to a given local environment. The implicit inclusion of the electronic screening effects in the model makes it fully compatible with non-polarizable force fields, such as AMBER, CHARMM, OPLS etc, yet the model is essentially as computationally efficient as the original TIP3P or similar non-polarizable water models.

## 2. Theory

2.1 Electronic screening, MDEC model

One may ask if is it possible at all to describe interactions between real molecules, which are inherently polarizable, with a non-polarizable force field? In general, the answer to this question is negative; i.e. it is not possible for example to describe a small cluster of water molecules in vacuum, and water molecules in the bulk state with the same set of fixed charges, because in vacuum, an isolated water molecule has a dipole 1.85D, whereas in liquid bulk state the dipole is about 3D [17]. (The exact value of the liquid dipole is still debated; here and below, as previously [15-19], we rely on the results of most recent first principles simulations of liquid state water, refs. [20-21]). Thus different thermodynamics states or different environments require different sets of charges to reflect vastly different properties of water in such different conditions.

However, if simulations involve only configurations which are similar in some sense, such as in liquid water, there is a well-defined average molecular dipole moment or, more generally, a well-defined average charge distribution within the molecule. In this case, a typical set of fixed parameters, such as charges, can be introduced that reflects the averaged values [22]. The remaining relatively small fluctuations around the average can be included as an additional renormalization or scaling of the mean-field parameters [15-16]. The renormalization of the original mean-field charges is equivalent to a familiar screening by the electronic polarizable continuum with dielectric constant $\varepsilon_{el}$ (also known as $\varepsilon_\infty$), which is a measurable characteristic for a given condensed state. For example, for liquid water the electronic part of the dielectric constant $\varepsilon_{el}$ is 1.78, and for most organic materials this value is close to 2.0. The resulting theory, which we call MDEC – Molecular Dynamics in Electronic Continuum [15-16] – involves only effective scaled charges, and does not include electronic polarizability explicitly; however, the electronic polarizable continuum is part of the model, and has to be included in certain cases, such as calculation of free energy, or dielectric properties of the material, or vaporization energy (when a molecule is transferred from a condensed phase to vacuum).

As we discussed recently, many empirical non-polarizable force fields are in effect MDEC models [15-16], see also [23]. For example TIP3P or SPC/E models of liquid water are such models. The effective dipole moment in such models (~2.3D) is very different both from the vacuum value 1.85D, and from most of estimates of the actual dipole moment in the liquid state (~3.0D [20-21]). The additional re-normalization or scaling due to electronic screening of the liquid state moment, results in the effective value of the dipole ~2.3D. The effective parameters of non-polarizable empirical force fields such as AMBER[4-5], CHARMM[6], GROMOS[7], OPLS[8-9] etc., can also be understood within the MDEC framework.

The significant change of the dipole and overall polarization of a molecule in a condensed phase, compared with a vacuum, is due to two factors. First, in condensed phase a molecule is surrounded by a polarizable electronic continuum with dielectric constant $\varepsilon_{el}$. The reaction field of the polarized medium surrounding a given molecule results in its re-polarization. In addition, there is a local environment of charges, such as those of four hydrogen bonded neighbors in liquid state water, which results in additional re-polarization of the molecule. Given that electronic dielectric property $\varepsilon_{el}$ is about the same in all materials, in the bulk water and inside a protein, for example, different local environments are mainly differ in local charge distribution. The idea of the following treatment is to develop a polarizable model that would implicitly contain the effect of the electronic polarizable continuum, and at the same time would react explicitly to local charge environment, thus separating the two effects.

2.2 MDEC Hamiltonian

The effective Hamiltonian of the MDEC model can be derived explicitly in a model of polarizable point charges [15-16], or polarizable point dipoles (see SI). However, since the final answer is quite simple, it is easier to refer to an intuitive picture rather than to a formal derivation, and write down such a Hamiltonian from the start. The illustration is shown in the Fig.1.

The effective Hamiltonian can be constructed as follows. We evaluate the energy of assembling a given configuration from separated molecules. In the first step, a water molecule is transformed from a gas-phase state to that in the condensed phase; i.e. the molecule is re-polarized ($\mu_0 \rightarrow \mu$), and generally modifies its state (defined by geometry, partial charges, etc) to that corresponding to the condensed phase. This step gives rise to self-energy of re-polarization $E^{self}(\mu)$, which is equivalent to Berendsen's additional term of SPC/E model [24], $E^{self}(\mu) = \dfrac{(\mu-\mu_0)^2}{2\alpha}$, but can also include the strain energy due to change of the molecular geometry and other quantum corrections [25].

In the next step-2, the molecule is transferred from vacuum to the electronic continuum. The energy cost of such a process is solvation energy $\Delta G^{el}(\varepsilon_{el}, q_i)$, i.e. solvation of the molecule with the liquid state parameters in the dielectric medium of $\varepsilon = \varepsilon_{el}$. This solvation energy, as well as the self-energy term, obviously do not depend on the configuration of the system, and considered to be constants in fixed charge models.

In the final step-3, the interactions with other molecules are added; these interactions occur in the polarizable electronic continuum, therefore all electrostatic interactions are screened, or reduced, by a factor $1/D(\varepsilon_{el})$ which depends on the dielectric constant $\varepsilon_{el}$, and on the electrostatic model of the molecule.

Thus, the electrostatic part of Hamiltonian in MDEC model has the following form:

$$U^{elec}(r_1,...,r_N) = \sum_{m=1}^{M} E_m^{self}(\mu_m) + \sum_{m=1}^{M} \Delta G_m^{el}(\varepsilon_{el}, q_i) + \frac{1}{2} \sum_{j \neq i}^{N} \frac{1}{D_{ij}(\varepsilon_{el})} \frac{q_i q_j}{r_{ij}} \qquad (2.1)$$

Here the last summation is over charges of different molecules, and the atomic indices at $D(\varepsilon_{el})$ indicate that the screening factor depends on the type of molecules to which the interacting pair of atoms belong.

In the simplest model of point charges, such as in TIP3P or similar, the screening factor $1/D(\varepsilon_{el}) = 1/\varepsilon_{el}$; however, in some other models the factor can be different [15]. (For example, for polarizable point dipoles [26-27], the screening factor is $\frac{1}{\varepsilon_{el}}\left(\frac{\varepsilon_{el}+2}{3}\right)^2$.)

The treatment of the constant solvation energy $\Delta G^{el}(\varepsilon_{el}, q_i)$, requires consideration of the molecular cavity which reflects molecular size and shape. This cavity is schematically shown in Fig. 1. The definition of the molecular cavity has been the subject of many discussions [26-27] and the computation of the electric field acting on a molecule in polarizable medium has always been one of the major problems in the continuum theories of electric polarization [26]. It should be stressed that the molecular cavity *does not* appear in the interaction terms, at least in the model of point charges.

The screened interactions for the point charge model ($D = \varepsilon_{el}$) can be described by the effective (screened) charges $q^{eff}$:

$$q_i^{eff} = q_i/\sqrt{\varepsilon_{el}}, \tag{2.2}$$

with the same scaling of the effective dipole of the molecule. Note the scaling does not break charge neutrality of the system because all charges are screened by the same factor. In these notations the electrostatic part of the MDEC model Hamiltonian is written as

$$U^{elec}(r_1,...,r_N) = \sum_{m=1}^{M} E_m^{self}(\mu_m) + \sum_{m=1}^{M} \Delta G_m^{el}(\varepsilon_{el}, q_i) + \frac{1}{2}\sum_{j \neq i}^{N} \frac{q_i^{eff} q_j^{eff}}{r_{ij}}. \tag{2.3}$$

The first two terms do not explicitly contribute to intermolecular forces $F_i = -\frac{\partial U}{\partial r_i}$; however, they need to be included in the potential energy, enthalpy and especially in solvation energy calculations [15,17-18,28-30]. The appearance in theory of these new terms also suggests a new approach for the parameterization of bonded potential. Indeed, since $\Delta G^{el}(\varepsilon_{el}, q_i)$ depends on the molecular geometry the bond, angle and torsion energy terms, parameterized in common force fields[4-9] in vacuum, should be corrected by the changes in the electronic solvation energy relevant to the condensed phase.

*Relation to Standard force fields models*

The above form of electrostatic interactions (last term in eq. (2.3)) is the same as that in conventional non-polarizable force fields of AMBER[4-5], CHARMM[6], GROMOS[7] or OPLS[9], where the atomic partial charges can be understood, at least approximately, as "scaled MDEC charges" [15-16]. However, this is only true for neutral groups.

In contrast, the ions, and charged groups, are described in standard non-polarizable force fields, such as CHARMM or AMBER, by their original integer charges (e.g. $\pm 1$, for Na$^+$ and Cl$^-$) completely disregarding the effect of electronic dielectric screening $1/D(\varepsilon_{el})$ inherent to the condensed phase medium. The direct interaction of such bare charges is obviously overestimated by a factor of about 2, $\varepsilon_{el}$, and the strength of ion-solvent interaction is overestimated by a factor about 0.7, $\sqrt{\varepsilon_{el}}$ ; see relevant discussion in Refs. [15-16,31-33].

Summarizing, the MDEC force field is similar but not entirely equivalent to standard non-polarizable force fields; the exception is for ionic groups. To make these force fields to be uniformly consistent with the idea of electronic screening, an additional scaling of ionic charges should be introduced [15-16].

*Water models*

Similarly, TIP3P[34] or SPC/E[24], and similar models of water, involve empirical charges that can be understood as scaled charges. Thus, the dipole moment of a water molecule in vacuum is 1.85D; in liquid state, however, the four hydrogen bonds to which each water molecule is exposed on average strongly polarize the molecule and its dipole moment becomes somewhere in the range of 2.7D to 3.2D [20-21,35]. [As we already mentioned, since in ab initio simulations of bulk water the water dipole cannot be defined unambiguously and depend, in general, on the partitioning scheme used [36], the exact value of the liquid dipole remains a matter of debate. Nevertheless, most of the studies[20-point to the average moment value which is more or less 1 Debye higher than that in the gas phase. Here we rely upon ab initio simulations and the extensively validated partitioning scheme of refs [20-21].] The significant increase of the dipole from $\mu_0 = 1.85D$ to a value $\mu \approx 3D$[20], or even larger[21], can be easily demonstrated [15-17] by the Kirkwood-Onsager model [37-38], which estimates the enhanced polarization of a molecule due to the reaction field of the polarized environment. Yet, the dipole moment of the TIP3P water model is only 2.35D. The specific value of the TIP3P dipole can be understood as a scaled value, so that the dipole-dipole interactions are screened in the electronic continuum by a factor $1/D(\varepsilon_{el})$. The scaling factor for neutral polar molecules, unfortunately, cannot be determined precisely. However, reasonable value for a model of point charges should be close to $1/\sqrt{\varepsilon_{el}}$, i.e. about 0.75 (for water $\varepsilon_{el}$=1.78), which results in the value of effective dipole[16-18] $\mu^{eff} \simeq 2.35D$.

The scaled nature of charges of TIP3P water model becomes critically important when interaction with a solute is considered. For example, if the charge of say Na$^+$ ion is assigned to be +1, then it is obviously inconsistent with the charges of water model; as the latter are scaled, while the charge of the ion is not. Clearly the strength of interaction is overestimated in this case by a factor $1/\sqrt{\varepsilon_{el}}$.

A detailed discussion of how the non-polarizable models, such as TIP3P and SPC/E, capture the polarizability effects in typical polar environment is given in our recent paper [17]. In biological applications, however, the solvation conditions can be variable across the studied object and essentially different from the reference environment for which the parameters were empirically chosen. A significant dependence of the water dipole moment on the polarity of the environment was shown to appear [15] in the media with dielectric constant $\varepsilon < 20$. The dipole moment of a water molecule in such media is significantly lower than the value $\mu_l$ of liquid water. Obviously, in low-dielectric environments, such as proteins or membranes, water should be modeled using potentials different than those of TIP3P or SPC/E. Thus, one cannot avoid the need for an effective mechanism for adjusting molecular parameters to different solvation conditions. Such a model built within the MDEC framework is discussed next.

2.3 Mean Field Polarizable Model of Water

A linear relation between equilibrium values of dipole and acting fields is obtained by the variational principle $\delta H / \delta \mu_m = 0$, provided by a quadratic form of the Hamiltonian (2.1) over moments $\mu_m$ in the dipolar approximation: $E_m^{self} = \dfrac{(\mu_m - \mu_0)^2}{2\alpha}$ and $\Delta G_m^{el} = -\dfrac{1}{2} \sum_i K_{ij}^m(\varepsilon_{el}) q_i^m q_j^m \approx -\dfrac{1}{2} f(\varepsilon_{el}) \cdot \mu_m^2$. Thus, in electronic continuum, as in vacuum, there is a simple linear relationship between the effective dipole $\mu_m^{eff}$ of the molecule, and the effective electric field $E_m^{eff}$ acting on it.

$$\mu_m^{eff} = \mu_{EC}^{eff} + \alpha^{eff} \cdot E_m^{eff} \qquad (2.4)$$

The corresponding polarizability $\alpha^{eff}$ is also an effective quantity, which is not the same as vacuum polarizability $\alpha$ of the molecule. Here $\mu_{EC}^{eff}$ is the dipole moment in electronic continuum, i.e. the moment in hydrophobic environment with no charges around. Given that the electronic continuum is about the same for all media, different environments then would be different only by the presence of the last term. An averaged value of the dipole corresponds to the average field of the surrounding charges acting on the molecule. The above relation thus allows one to adjust the effective parameters to the local charge environment.

In the following, we implement the above principles in a polarizable water model. The model is optimized for biological applications with existing force fields such as AMBER and CHARMM. With this aim the parameters of the new model were selected to reproduce the properties of the TIP3P water rather than the properties of liquid water. Although it is recognized that the TIP3P is not the best model for water simulation, the parameters of the force fields such as AMBER or CHARMM were optimized for use with this specific potential. Thus, in biological applications the use of a better solvent model does not guarantee compatibility or improvement of the protein-solvent interactions.

The first parameter $\mu_{EC}^{eff}$ (effective equilibrium moment of polarizable TIP3P molecule in the electronic continuum of $\varepsilon_{el}$) can be estimated computationally by the standard methods of continuum electrostatics [39] or even analytically assuming the spherical shape of the molecular cavity. For example, for a point dipole in a spherical cavity, one can find the following expression [15,17]:

$$\mu_{EC}^{eff} = \frac{(\mu_0/\sqrt{\varepsilon_{el}})}{1 - \frac{6}{\pi}\frac{(\varepsilon_{el}-1)}{\varepsilon_{el}+2}\frac{2(\varepsilon_{el}-1)}{2\varepsilon_{el}+1}}, \quad (2.5)$$

where $\mu_0$ (1.855D) is the gas value of water dipole. For water $\varepsilon_{el}$=1.78, and the above expression gives a value close to 1.6D. However, as we already mentioned, the point dipole model is not an accurate substitute for charge distribution of TIP3P. Therefore a more accurate model based on PCM calculations [39] was actually used. The obtained value of $\mu_{EC}^{eff}$ is 1.8 D, which is close to the gas phase dipole moment 1.855D, but here it has an absolutely different meaning. (The reason for the close value is that in electronic continuum, the gas-phase dipole increases, but the electronic screening decreases it's effective value. The action of these two factors may compensate each other, as in this case.)

The effective polarizability $\alpha^{eff}$ then is directly obtained from the original TIP3P model, using $\mu_{EC}^{eff}$=1.8D, the bulk value of the effective TIP3P dipole, $\mu_{TIP3P}^{eff}$=2.35D, and the average effective field acting on a water molecule in bulk in TIP3P model,

$$\alpha_{TIP3P}^{eff} = \left(\mu_{TIP3P}^{eff} - \mu_{EC}^{eff}\right)\Big/\left\langle E_Z^{eff} \right\rangle_{TIP3P} \tag{2.6}$$

here $\mu_{TIP3P}^{eff}$ is the molecular moment in TIP3P model, $\left\langle E_Z^{eff} \right\rangle_{TIP3P}$ is average effective field acting on TIP3P water molecule in the bulk, see Fig. 2.)

The field acting on the water molecule was taken at the position of oxygen, or, alternatively, as weight-averaged over all charged sites of the molecule. In the following, we discuss only the former model, because no improvement from the weight-averaging was observed in the test simulations. The dynamical changes of molecular moment $\mu_m^{eff}$, in response to the external field, are modeled by modifying the atomic partial charges, which in TIP3P fixed molecular geometry are linearly related to the dipole moment.

*The averaging technique and self-consistency*

To reflect the averaged or mean-field nature of the local polarization environment, the field acting on the molecule was not taken to be instantaneous as in the usual polarizable models, but instead as a temporally averaged one, using the following damping technique.

$$\mu^{eff}(t) = \mu_{EC}^{eff} + \frac{\alpha^{eff}}{\tau}\int_0^t dt' E^{eff}(t-t')e^{-t'/\tau} \qquad (2.7)$$

The physical meaning of the relation can be recognized by considering limiting cases. For example, for a constant field $E^{eff}(t)=E_0$ and averaging time longer than the relaxation, the relation is reduced to the traditional law: $\mu(t) = \mu_{EC}^{eff} + \alpha^{eff} \cdot E_0$. If the relaxation time $\tau$ is such that it is much shorter than characteristic time-scale of the field $E^{eff}(t)$, which for liquid water is around 1ps, the above expression is equivalent to instantaneous field; in the opposite limit it is equivalent to time-averaged field. In the intermediate regime, the external field would be averaged over finite time $\tau$, reflecting the possible change of the local environment of a given molecule.

In practice, the implementation of the temporal averaging uses a discrete differential form of the relation (2.7):

$$\left(\mu^{eff}\right)_k = \left(\mu^{eff}\right)_{k-1} + \frac{\mu_{EC}^{eff} + \alpha^{eff} \cdot \left(E^{eff}\right)_k - \left(\mu^{eff}\right)_{k-1}}{(\tau/\Delta t)} \qquad (2.8)$$

where $\left(E^{eff}\right)_k$ is the instant field on the current step $k$, $\left(\mu^{eff}\right)_k$ and $\left(\mu^{eff}\right)_{k-1}$ are the dipole moments on the current and previous step, respectively, and $\Delta t$ is the simulation time step. The field and the dipole moments correspond to a given particle and reflect the character of local environment. In eq.(2.8) the damping factor $\frac{1}{(\tau/\Delta t)}$ ensures that the $k$-th instant moment value $\mu_{EC}^{eff} + \alpha^{eff} \cdot \left(E^{eff}\right)_k$ will be reached not immediately but approximately in $N_{damp}=(\tau/\Delta t)$ steps. Thus, significant stochastic fluctuations of the instant field are damped on each time step by the factor $1/N_{damp}$, effectively resulting in adjustment of the moment to the average field (2.7).

For accurate averaging, the relaxation time $\tau$ is chosen to be sufficiently large ($N_{damp}=\tau/\Delta t \gg 1$) to provide a smooth dipole moment adjustment (2.8) on each time step. The adiabatic variation of moments also allows achieving self-consistency of all the water molecules in the system.

The dipole moment variation is achieved by adjusting the partial charges of water to reproduce the moment value (2.8). It is well known that this way of describing polarizability, like all other models of the fluctuating charge type[40], has a deficiency of handling the out-of-plane water polarization. Nevertheless, the approach allows straightforward implementation of the averaging procedure (2.7), and is computationally more efficient due to no additional interacting sites is required, unlike the drude oscillator[41-42] and point multipole techniques.

It should be noted that since the charge adjustment procedure (2.8) is just linear in the number of particles (similarly to the velocity scaling in the temperature coupling procedure), and no extra computations of forces is required for the self-consistency of polarization, the computational efficiency of the model is almost the same as that for the non-polarizable TIP3P or SPC/E type of models. Namely, the performance observed in the benchmark simulation of box of 2048 polarizable water molecules is within 10% of that for the original TIP3P.

*Non-linear polarization*

Finally, to prevent the "polarization catastrophe", a typical problem for all linear polarizable models, we make the polarizability $\alpha^{eff}$ electric field-dependent for large field strengths as discussed in ref [42]:

$$\alpha^{eff}(E^{eff}) = \begin{cases} \alpha^{eff} & , \quad E^{eff} \leq E^* \\ \alpha^{eff}\left(E^*/E^{eff}\right)\left(2 - E^*/E^{eff}\right), & E^{eff} > E^* \end{cases} \quad (2.9)$$

here $E^{eff}$ is the instant effective field used for damping on each time step while the critical field $E^*$ is an adjustable parameter of the model.

### 3. Simulation details and Results

The mean field polarizable TIP3P model (MFP/TIP3P) was implemented into Gromacs [43] MD simulation package ver 4.0.7. Parameters of the model are listed in Table 1, along with the corresponding values for the non-polarizable TIP3P model [4-5]. The partial charges in the model are dynamically changed as described above; the geometry and van-der-Waals parameters were taken from the TIP3P model [4-5] without modifications; the values of the parameter $\mu_{EC}^{eff}$ (equilibrium moment in the electronic continuum of $\varepsilon_{el}$) and effective polarizability $\alpha^{eff}$ were calculated as discussed in the previous section. The one adjustable parameter of the model $E^*$, the non-linear cutoff, was obtained by matching the density $\rho$ obtained in the bulk simulations with the density of TIP3P model.

The averaging time $\tau=5$ ps was used in the discussed simulations; the value was found as a compromise between accuracy and temporal resolution of the model. The critical parameter basically defines how fast the model responds to the changes in polar environment. The mean-field model resolution is physically limited by the timescale of hydrogen bond lifetime ($\tau_H$). In liquid water $\tau_H$ is about 1 ps, while several instances of the bond making and breaking are needed for the accurate averaging. The dependence of the precision of simulated water properties on the parameter $\tau$ can be found in SI.

The seemingly large averaging time should not be critical for most of biological applications. For example, in uniform media such as the bulk water, even though molecules can diffuse for 5 ps to the distance about 5 Å ($\sqrt{6\tau D}$, here $D$ is a diffusion coefficient), the polar environment is about the same at any location. In non-uniform protein media the water migration is typically slowed down by the strong hydrogen bonding or steric interactions. If the environment significantly changes within the diffusion distance, which is the case for interfacial systems, the mean-field model still will perform better than the original TIP3P.

*Simulation protocol*

The liquid water properties were obtained in MD simulations of cubic cell formed by 2048 water molecules, except for the ion solvation energies which were modeled in a box of 1000 water molecules. The protein simulations were carried out with Amber force field[4-5] scaling the charges of ionic groups as discussed in the text. Since inside the protein the water motion is naturally restrained no artificial restrains were applied to achieve the initial equilibration of solvent polarization with the temporally averaged protein field. The electrostatic interactions were treated by the PME technique with a real space cutoff of 12Å and sixth order spline for mesh interpolation. The van der Waals interaction cutoff of 12 Å was used along with the long-range dispersion corrections for pressure and energy. Nonbonded pair list was updated each 11 fs. The Nose-Hoover thermostat with the coupling constant 0.1 ps and Parrinello-Rahman barostat with the coupling constant 0.3 ps were applied to keep the temperature at 298 K and pressure 1 atm. To maintain a correct temperature it was essential to remove the center of mass motion on every time step. For each modeled property, the statistics was collected over 4 to 8 MD simulations started from different configurations. For each simulation the system was equilibrated first during a 0.5ns run, followed by 10ns data collection run integrated with a 1 fs time step.

*Liquid water simulations*

As shown in Table 2, the MFP/TIP3P model accurately reproduces the liquid properties obtained in TIP3P simulations at ambient conditions. Radial distribution functions $g_{OO}$, $g_{OH}$ and $g_{HH}$ are exactly the same for both models and not shown here. Some imperfection in the comparison of calculated properties with experimental data is not critical, because our goal is to reproduce TIP3P model in the bulk, disregarding its own imperfections.

Fig. 2 shows a comparison of the average field acting on the molecule in TIP3P and in MFP model.

The self-consistent nature of the model is demonstrated in Fig. 3, where the bulk water is simulated with arbitrary values of the molecular dipoles at the starting point, t=0. Independently of the initial charges, the ensemble finds self-consistent parameters that are exactly matching those of TIP3P; after initial relaxation, the parameters do not change, and the model is equivalent to the standard TIP3P model. The relaxation time of the ensemble depends on the size of simulated system. For a small cluster, the relaxation is expected to be close to $\tau=5$ ps. The steady-state values of the average dipole moment, electric field acting on the molecule, and the potential energy are compared with corresponding TIP3P values. To achieve an agreement with the energy of TIP3P model, it was essential to include the self-polarization energy $E^{self}$ and electronic solvation energy $\Delta G^{el}$ introduced by the MDEC (2.3); these two terms are usually missing in the non-polarizable force field models.

As shown earlier[16], the corrections (2.3) are also essential for the free energy calculation in low-dielectric environments[16] such as alkanes, non-polar nanopores, lipid membranes or hydrophobic protein interiors. In the high-dielectric environment, however, the missing terms are luckily compensated[16,28] by the empirical choice of ionic radii[4-5]. Thus, to exclude the solvent independent component $\Delta G^{el}$ from the comparison the hydration free energies in Table 2 were simulated in the standard way [without MDEC corrections (2.3)].

The MFP/TIP3P data for hydration free energies of $Cl^-$ and $Na^+$ ions and structure of their solvation shells (see Fig. 4), were found to be practically the same as in TIP3P model.

The reason is that ions with charge $\pm 1$ induce the local electric field and solvent polarization which are similar to that in the bulk water. Interestingly, both charge options for anions: the bare value $-1e$ and scaled value $-0.7e$ are not much different from that of water oxygen ($-0.834e$).

We wish to stress that of importance here is not the agreement of calculated free energies with experimental data because in high-dielectric media such as water the quality of the results is determined mainly by a consistent choice of the ionic radii[4-5,48] rather than by the quality of solvent potential. There is also a number of subtleties resulting in ambiguity of both simulated and experimental values of hydration energies[31,48]. Our results instead aim to demonstrate here only an ability of the polarizable potential to reproduce the solvation properties of the original TIP3P.

Thus, in liquid water simulations the polarizable MFP/TIP3P model accurately reproduces the statistical ensemble and thermodynamic properties of non-polarizable solvent model and, hence, can substitute for the TIP3P model without introducing an inconsistency to the entire force field. Yet, in different solvation conditions, the polarizable model is capable to adjust its effective parameters to reflect the changes in the local polar environment which naturally should improve the electrostatic interactions in the condensed phase.

One additional important point for liquid water simulations should be stressed finally. The usual polarizable models are typically evaluated by considering situations that are very different from the bulk conditions, for example, by modeling water clusters in vacuum. Our mean field model, however, is not applicable for systems in vacuum because the model is essentially based on the assumption of the electronic continuum filling the entire space which is inherent for the condensed matter.

*Protein simulations*

To demonstrate the capability of the model to reflect vastly different conditions in a protein interior from those in liquid water, we modeled solvated Cytochrome *c* Oxidase (C*c*O), the terminal enzyme in the electron transport chain that included a large number of experimentally observed water molecules inside the protein. The biological function of CcO (reduction of oxygen to water and proton pumping) is essentially determined by the internal water [49-51]. The different conditions that exist inside the protein, in particular around the catalytic center of the enzyme, and the response of the internal water polarization to these conditions, are shown in Fig. 5. Different water molecules inside and outside the protein are identified by their location with respect to the $Cu_B$ center of the catalytic center of the enzyme, which is located right in the middle of the protein. The distance over around 50Å in Fig. 5 corresponds to outside bulk water surrounding the protein.

As seen in Fig. 5 (black line), the water polarization on average is lower in the protein interior (roughly, $r<35$Å) than in the protein exterior (35Å$<r<$50Å) and in the bulk solvent ($r>$50Å), reflecting the fact that the medium in the protein interior is typically less polar than in the protein exterior or in the bulk solvent. Note, the partitioning of the non-spherical protein onto the protein interior, exterior and solvent is approximate; in fact, all the regions at $r>20$Å include in different proportion the water molecules from the outside solvent. Even though the variation of the dipole moment values is mainly determined by the stochastic fluctuations, the difference between the regions is larger than amplitude of these fluctuations. It is seen that the individual water molecules inside the protein can have significantly different dipole moment than in the bulk. Obviously, the internal water cannot be always correctly modeled by the fixed charge TIP3P model ($\mu^{eff} \approx 2.35$D) commonly used in simulations.

In contrast, MFP/TIP3P model - for essentially the same computational cost - captures the variation of the local protein environment and corresponding changes in the water dipole moment. As seen in Fig. 5, the strongest polarization field is acting on the three water molecules ligated to $Mg^{2+}$ center of the enzyme, data for ($r$~12Å), those molecules have highest moments $\mu^{eff}$~2.8D. The lowest polarization field is experienced by water in the hydrophobic cavities of the enzyme. In such regions, apparently, water forms much weaker hydrogen bonds with the protein environment, compared to the bulk solvent, resulting in depolarization of the water dipoles down to $\mu_{EC}^{eff}$=1.8D. The later value corresponds to the solvation in non-polar hydrophobic media, which is described in MDEC model by an electronic continuum without any charges.

Although water cannot permanently reside in hydrophobic cavities, it can be temporarily trapped there in a meta-stable state, providing unique functionality to the enzyme, e.g. serving as a medium for proton transfer. The correct simulation of such meta-stable waters inside proteins is obviously of great importance. For example, in hydrophobic catalytic cavity of CcO water is produced by the reduction of oxygen; this water appears to play a key role in the proton pumping mechanism of the enzyme [49-51]. In CcO simulations, typically four water molecules are present in the catalytic cavity. As seen in Fig. 5, the dipole moments of these four molecules ($\mu^{eff}$ ~2.05D) are significantly lower than those in the bulk ($\mu^{eff}$ ~2.35D) or standard TIP3P, which reflects the low polarity of hydrophobic local environment of that part of the protein.

Thus, the mean field polarizable model adequately captures the induced polarization of water in the non-uniform protein environment. Although it is recognized that for realistic description the polarizability of the protein itself missing in our simulations can be also important, the above modeling already demonstrates major changes occurring with the internal water due to variations in number and strength of hydrogen bonds in the local protein environment. Thus, the substitution of the non-polarizable TIP3P solvent model by the MFP/TIP3P model naturally improves the simulation of water in non-uniform biological environments, with essentially no loss in computational efficiency.

## 4. Conclusions

In the present study we aimed at developing a polarizable solvent model that is applicable in simulations with the existing non-polarizable biological force fields. In addressing the problem of incompatibility of the standard polarizable water potentials and non-polarizable force fields, two ideas were described and incorporated into a new mean-field polarizable model.

1. The effects of polarization are partitioned into two components: those related to the generics electronic screening of electrostatic interactions, characteristic to the condensed phase, and those related to additional adjustment of water charges to local polar environment. The first component is treated in the continuum approximation by using *effective*, i.e. scaled by the electronic dielectric factor charges. As previously shown[15-18] the charge scaling is consistent with the nature of empirical parameters of non-polarizable force fields. In this case the total polarization is naturally described by re-adjustment of the non-polarizable effective charges of water model from the bulk liquid state to a new environment, such as inside of the protein.

2. To further match the spirit of non-polarizable effective force fields, the new model adjusts the water effective dipole (or equivalent point charges) not to the instantaneous field, as is usually done in standard polarizable models, but rather to the mean-field acting on the molecule. The mean-field nature of the model makes it identical to TIP3P or SPC/E type of models in liquid bulk phase; whereas inside a protein, or at interfaces, it readjusts the effective parameters of the model to match different local environment. The model is essentially as computationally efficient as the original TIP3P or SPC/E type of models and by the construction is much more compatible with the rest of non-polarizable force-field than a straight-forward fully polarizable model.

The model assumes that simulation is conducted in a condensed phase environment, where the concept of electronic polarizable continuum is applicable (thus the model *is not* suitable e.g. for simulation of small water clusters in vacuum). In non-polar environment, such as lipid membrane, described with a standard non-polarizable force field the dipole moment of our water model is not 1.85D as it would be in a vacuum, or in a straight-forward polarizable water model, but one corresponding to a water molecule in a medium with a dielectric constant of about 2.0. Inside a protein, such as cytochrome c oxidase studied in this work, the effective dipole of water varies in the range 1.9D to 2.8D reflecting different conditions existing in the interior of the protein. Whereas outside the protein, in the bulk water, the model finds a self-consistent value of the dipole that coincides with that of TIP3P or equivalent SPC/E model.

The developed theory provides a relation between the actual molecular charges and effective parameters of the non-polarizable force fields, bridging the gap between polarizable and non-polarizable models. Based on this formalism a mechanism of adjusting effective parameters of the mean field model to the local polar environment was implemented in the GROMACS computer code.

To maximize compatibility with existing biological force fields such as Amber and Charmm, the mean field polarizable (MFP) model was optimized to accurately reproduce properties of the TIP3P solvent in the liquid bulk conditions rather than the properties of actual water. Thus, in typical solvation conditions the resulting MFP/TIP3P model can replace the non-polarizable TIP3P without introducing inconsistency with the entire force field of AMBER or CHARMM.

In different solvation conditions such as a non-uniform protein environment the mean-field polarizable model adequately captures the character of the local environment, different from that of bulk water. Although, the approach can be criticized for missing the out-of-plane water polarization and protein polarization effects, the present model captures the major changes in water properties due to the variation in local order of hydrogen bonding. Thus the replacement of the non-polarizable TIP3P solvent with the polarizable MFP/TIP3P model can naturally improve biological simulations without losing computational efficiency which is important for large scale biological simulations.

## Acknowledgments

This work has been supported in part by the NSF grant PHY 0646273, and NIH GM054052.

## Associated Content
**Supporting Information Available:** Derivation of the MDEC effective Hamiltonian from the explicit polarizable model, the dependence of water model properties on the time resolution of the mean-filed MFP/TIP3P model.
This material is available free of charge via the Internet at http://pubs.acs.org.

**Table 1** Parameters of the polarizable and non-polarizable TIP3P model

| | $d_{OH}$, Å | ∠HOH, ° | $\varepsilon_O$, kcal/mol | $\sigma_O$, Å | $q_O$, e | $q_H$, e | $\mu_0$, D[a] | $\alpha^{eff}$, Å$^3$ | $E^*$, e/nm$^2$ | $\tau$, ps |
|---|---|---|---|---|---|---|---|---|---|---|
| TIP3P | 0.9572 | 104.52 | 0.152 | 3.15076 | −0.834 | 0.417 | 2.347 | 0 | | |
| MFP/TIP3P | 0.9572 | 104.52 | 0.152 | 3.15076 | [b] | [b] | 1.8 | 1.044 | 13.47 | 5 |

[a] The permanent dipole moment which is $\mu^{eff}_{TIP3P}$ for TIP3P and $\mu^{eff}_{EC}$ for MFP/TIP3P.

[b] The charges in MFP/TIP3P model are adjustable to reproduce the dipole moment given by eq. (2.8).

**Table 2** Liquid water properties at ambient conditions [a)].

| Property | | exp | TIP3P | MFP/TIP3P |
|---|---|---|---|---|
| Average dipole moment | $<\mu>$, D | | 2.3470 | 2.3455(1) |
| Average electric field [b)] | $<E>$, e/nm$^2$ | | 10.906(0) | 10.906(1) |
| Vaporization energy | $\Delta u$, kcal/mol | −9.92 [c)] | −9.58(0) | −9.58 (0) |
| Density | $\rho$, g/cm$^3$ | 0.997[44] | 0.986 (0) | 0.986 (0) |
| Shear viscosity | $\eta$, mPa s | 0.89[44] | 0.32 (1) | 0.32 (1) |
| Diffusion coefficient [d)] | $D$, 10$^{-5}$ cm$^2$/s | 2.3[45] | 6.1 (1) | 6.1 (1) |
| Dielectric constant | $\varepsilon$ | 78.4[44] | 98 (2) | 100 (2) |
| Hydration free energy [e)] of Cl$^-$ | $\Delta G$(Cl$^-$), kcal/mol | −89.1[46] | −85.8 (1) | −85.7 (1) |
| Hydration free energy [e)] of Na$^+$ | $\Delta G$(Na$^+$), kcal/mol | −88.7[46] | −84.1 (1) | −84.5 (2) |

[a)] The numbers in parentheses indicate the statistical errors of the last digits (standard deviation of the mean estimated from block averages).
[b)] Electric field component along the molecular dipole at the position of oxygen atom.
[c)] The value $\Delta H_{vap}$ from ref [44] corrected by $RT$ term.
[d)] The calculated diffusion coefficients include the finite-size correction[47].
[e)] The calculated free energies do not include the finite-size correction[31] to be consistent with the approach used in the ion parameterization[4-5].

## Figure Legends

**Fig.1** Partitioning onto steps the bulk assembling process in MDEC model. In the step-1, the gas phase molecule with moment $\mu_0$ is polarized and transforms its geometry to its condensed phase state; next step-2 involves solvation in electronic continuum; and in the step-3 the screened molecular interactions are added as described in the text.

**Fig.2** Effective Field distribution in the bulk water. Triangles, circles and squares stand, respectively, for the X-, Y- and Z-component (in the local water basis) of the field obtained in simulations with TIP3P model. The solid line represents the corresponding distributions for MFP/TIP3P, model.

**Fig.3** Relaxation, and fluctuations of the average effective dipole moment, the field (in the local water basis), as well as average potential energy per molecule in NVT simulations of the bulk water at $T$=298K. The relaxation of the ensemble of 2048 MFP/TIP3P molecules started from almost zero partial charges ($\mu^{eff} \approx 0$) is shown on the left; fluctuations of the TIP3P ensemble are shown on the right for comparison.

**Fig.4** Solvation of ions in MFP/TIP3P model of water. Top – distribution of dipole moment of MFP/TIP3P water surrounding the ions. Middle – distribution of the instant electric field acting on the molecule. Bottom – radial distribution function (oxygen with respect to the ion). The distributions obtained with scaled and unscaled charge of ions are represented by red and blue solid lines, respectively. For the comparison the distributions of the TIP3P solvent around the ions with scaled charge are shown by black dashed line. The MFP (red) and TIP3P (black dashed) rdf curves are almost identical.

**Fig.5** Distribution of MFP/TIP3P dipole moment of water molecules solvating Cytochrome c Oxidase. The distribution snapshot corresponds to a single configuration taken after 2ns simulation. Individual water molecules are identified by their distance *r* from $Cu_B$ center of the enzyme, for convenience. Triangles, squares and circles refer to individual molecules that are ligated to $Mg^{2+}$, residing in catalytic cavity, and all other, respectively. The solid black line shows the average over a spherical slice of radius *r* averaged over configurations of the last 1ns. The average line is shown for distances at which the slice averaging is statistically meaningful. See text for details.

**Figure 1**

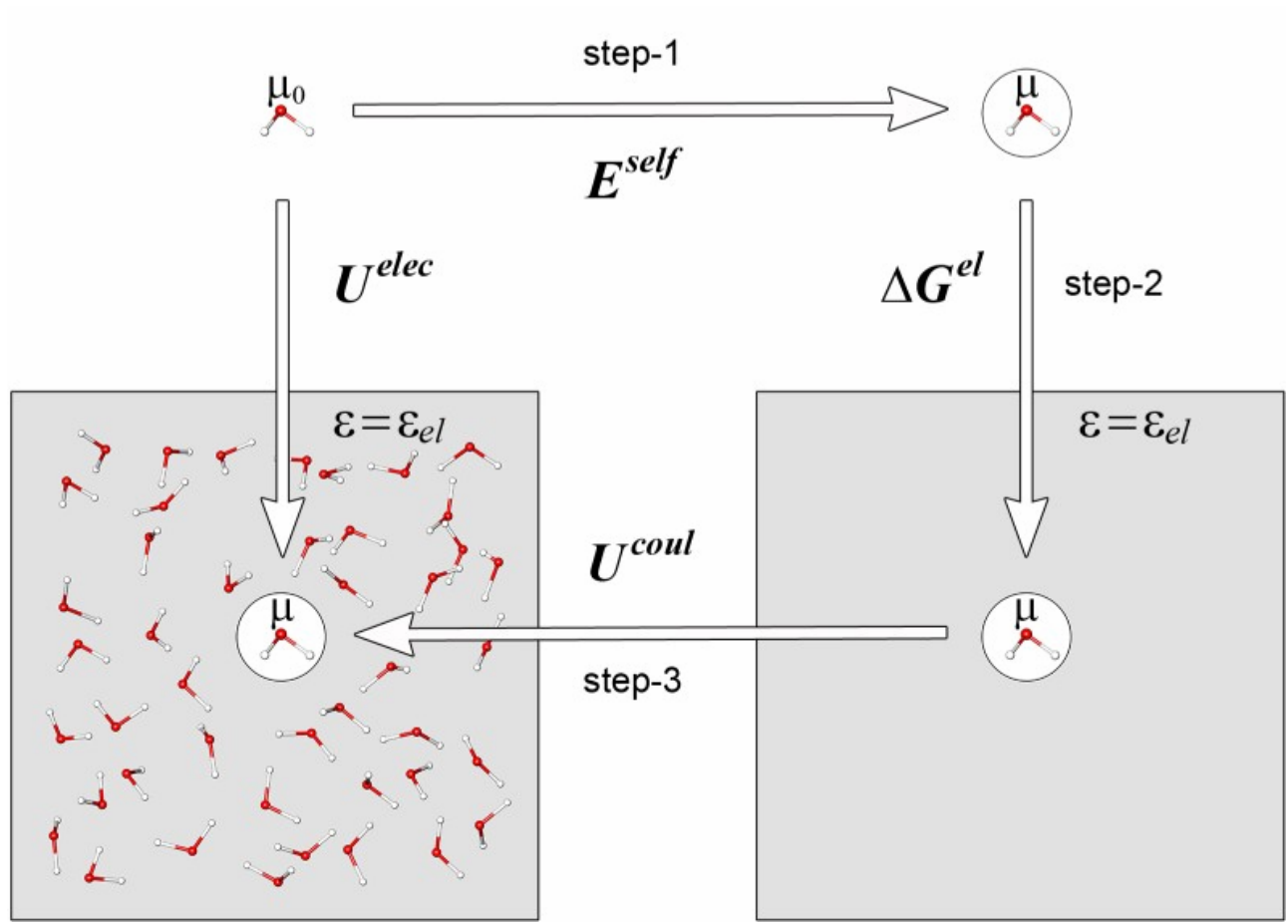

**Figure 2**

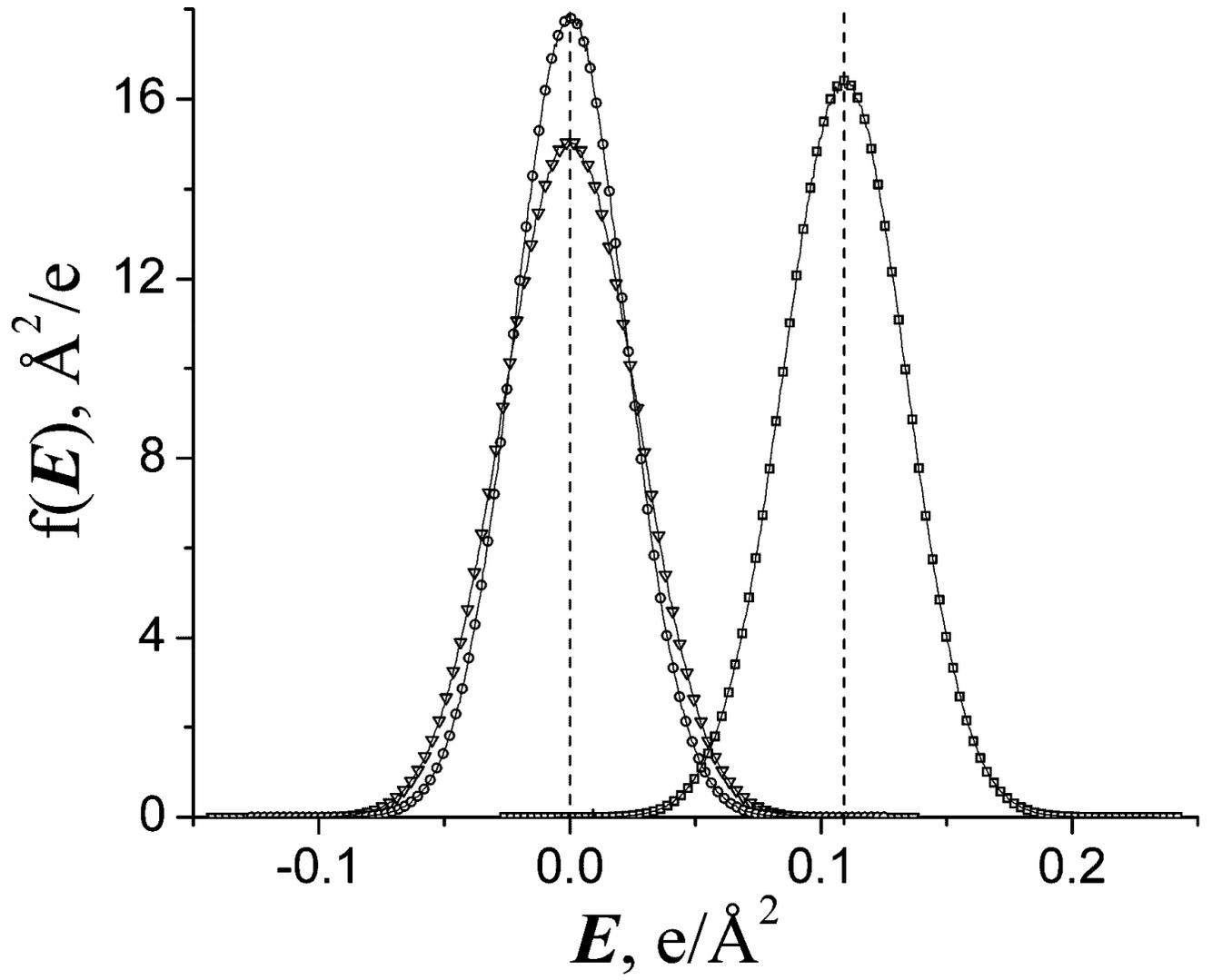

Figure 3

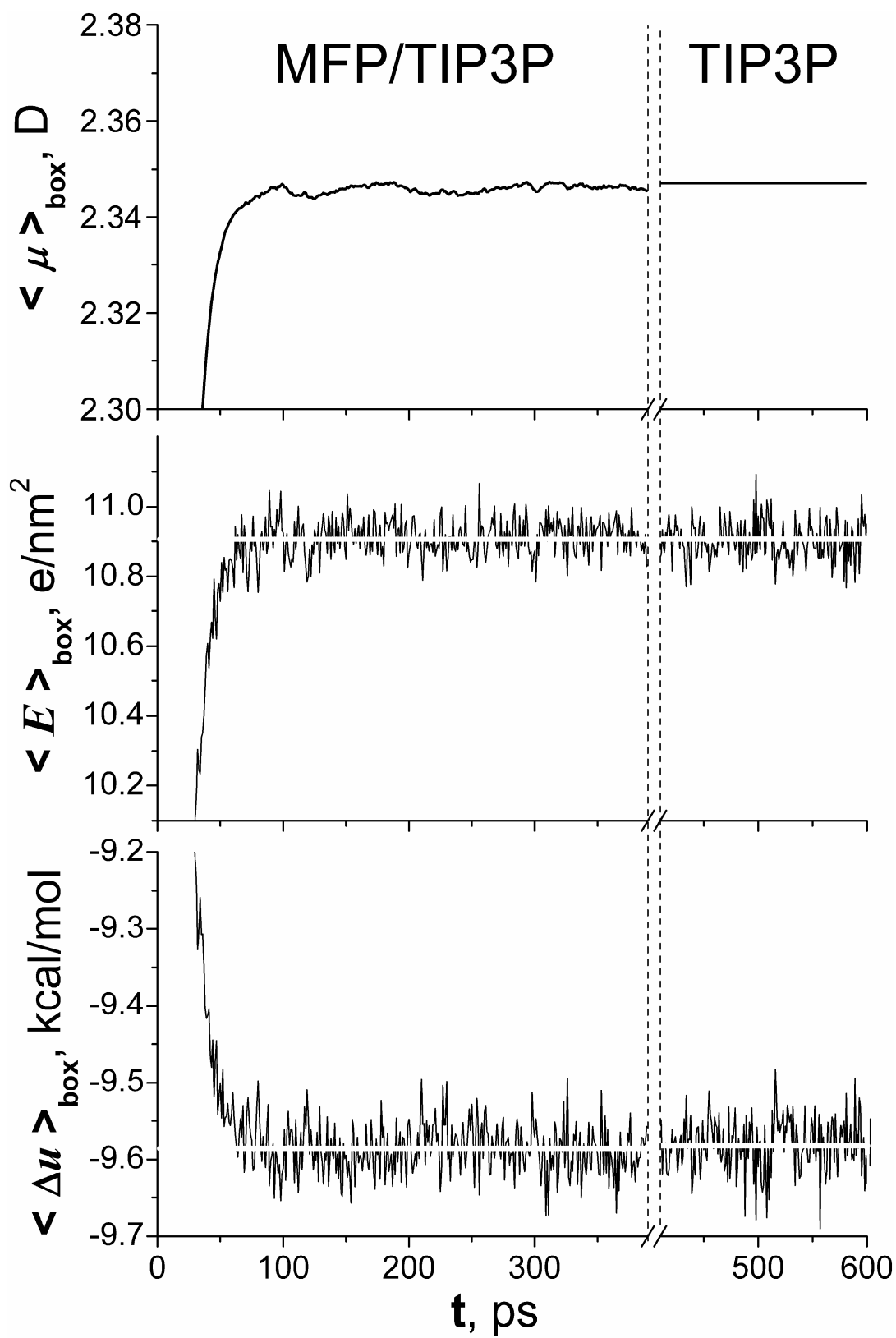

**Figure 4**

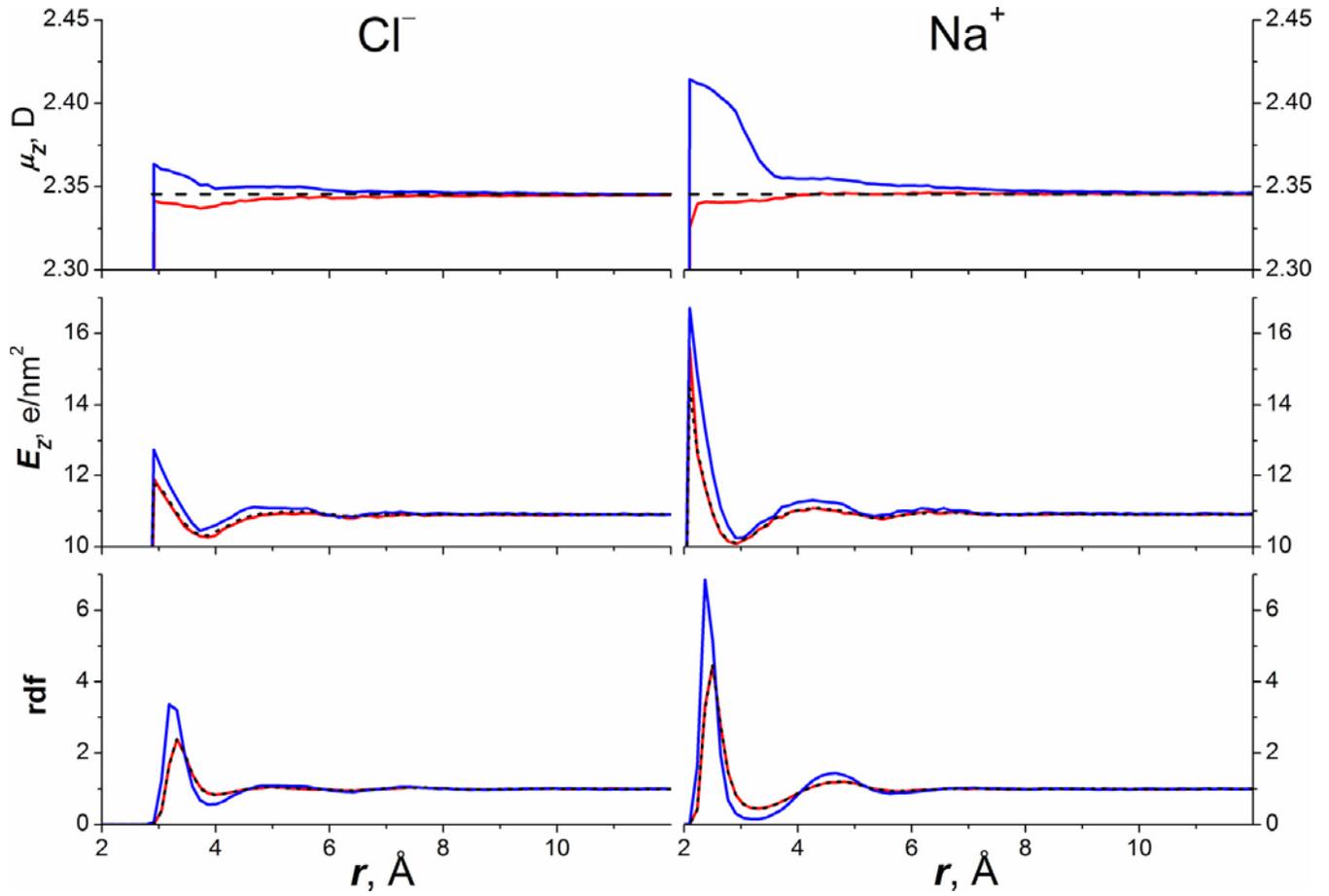

**Figure 5**

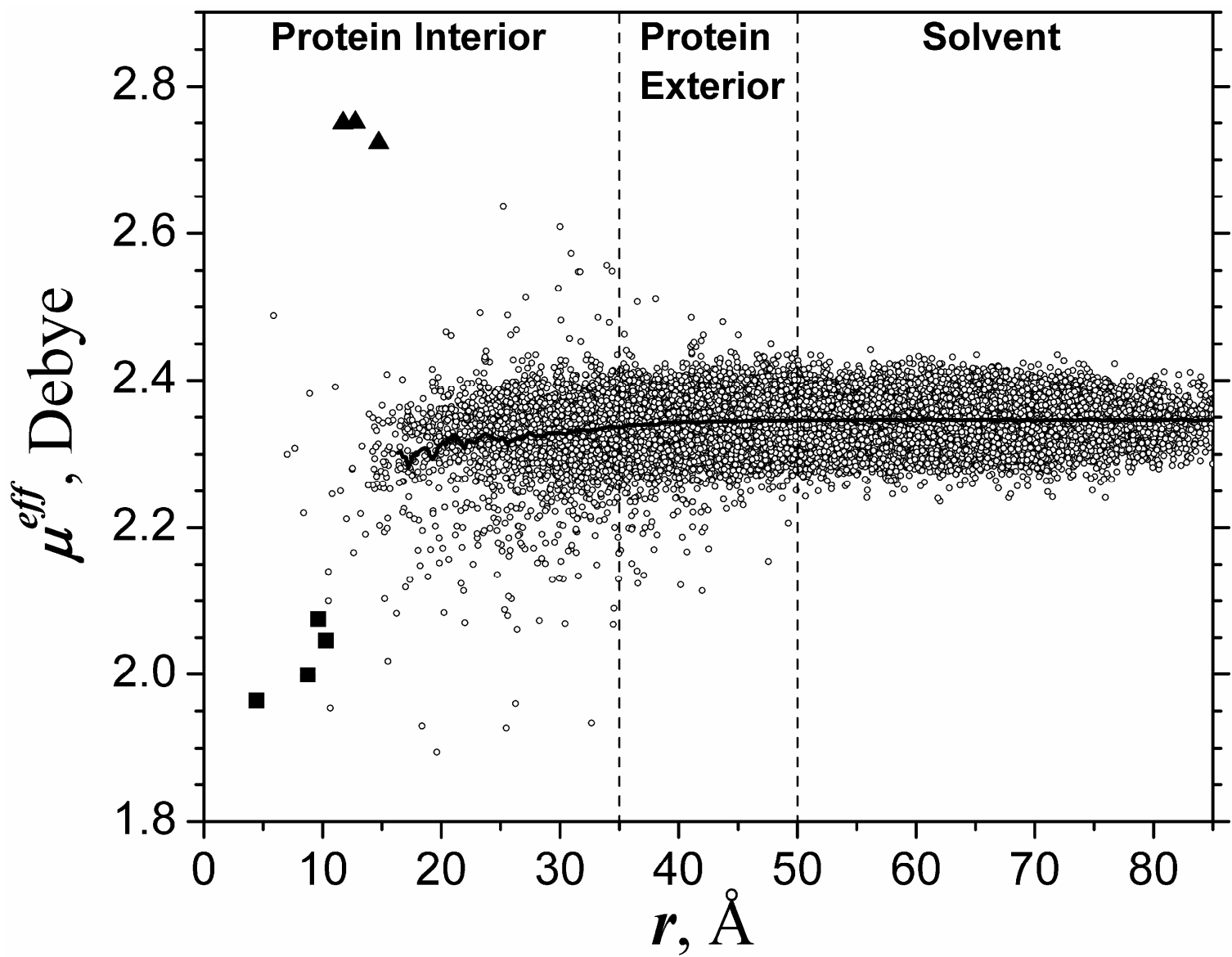